\documentclass[aps,pra,twocolumn, noshowpacs, groupedaddress,nopacs,floatfix]{revtex4-1}
\usepackage{graphicx}
\usepackage{amsmath}
\usepackage{amssymb}

\usepackage[font=small, labelfont=bf]{caption}

\def\<{\langle}
\def\>{\rangle}

\begin{document}
\title{Efficient fiber-optical interface for nanophotonic devices}

\author{T. G. Tiecke$^{1,2}$}
\thanks{These authors contributed equally to this work}
\author{K. P. Nayak$^{1,3}$}
\thanks{These authors contributed equally to this work}
\author{J. D. Thompson$^{1}$}
\thanks{These authors contributed equally to this work}
\author{T. Peyronel$^{1}$}
\author{N. P. de Leon$^{1,4}$}
\author{V. Vuleti\'{c}$^{2}$}
\author{M. D. Lukin$^{1}$}
\affiliation{$^{1}$Department of Physics, Harvard University, Cambridge MA 02138}
\affiliation{$^{2}$Department of Physics, MIT-Harvard Center for Ultracold Atoms, and Research Laboratory of Electronics, Massachusetts Institute of Technology, Cambridge, Massachusetts 02139}
\affiliation{$^{3}$Center for Photonic Innovations, The University of Electro-Communications, 1-5-1 Chofugaoka, Chofu, Japan,}
\affiliation{$^{4}$Department of Chemistry and Chemical Biology, Harvard University, Cambridge, MA 02138, USA}

\begin{abstract}
We demonstrate a method for efficient coupling of guided light from a single mode optical fiber to nanophotonic devices. Our approach makes use of single-sided conical tapered optical fibers that are evanescently coupled over the last $\sim10\,\mu $m to a nanophotonic waveguide. By means of adiabatic mode transfer using a properly chosen taper,  single-mode fiber-waveguide coupling efficiencies as high as $97(1)\%$ are achieved. Efficient coupling is obtained for a wide range of device geometries which are either singly-clamped on a chip or attached to the fiber, demonstrating a promising  approach for integrated nanophotonic circuits, quantum optical and nanoscale sensing applications.
\end{abstract}

\maketitle

\section{Introduction}
The field of nanophotonics \cite{joannopoulos08} opened new avenues for applications such as nanophotonic integrated circuits\cite{chen11,miller10}, 
sensing \cite{pohl84,lewis84,tan92,yan12} and scalable quantum information processing \cite{kimble08,ladd10,vanmeter10}. Moreover, sub-wavelength confinement of optical fields enabled strong light-matter interaction at the single quantum level \cite{yoshie04,tiecke14}. 
A major challenge in the field is to efficiently integrate the nanophotonic devices with conventional optical fiber networks. This challenge is due to a large mismatch between the size of the fundamental mode of the optical fiber and that of the optical modes of nanophotonic devices. This mismatch has to be bridged in order to achieve efficient coupling. 
Highly efficient coupling is crucial for applications such as quantum repeaters \cite{briegel98} or quantum networks \cite{kimble08} since the performance of these systems, in the limit of many nodes, deteriorates near-exponentially with photon loss between individual nodes. Additionally, highly efficient coupling enables distribution of non-classical states of light which are extremely fragile to photon loss.

A wide range of coupling techniques are currently  being explored, including grating coupling \cite{chen11} and end-firing from macroscopic fibers \cite{cohen13} where  coupling efficiencies up to 70-80\% to on-chip waveguides have been achieved. More recently, on-chip photonic waveguides have been coupled to the waist of a biconical fiber taper \cite{groblacher13} with an efficiency as high as 95\%.

In this Letter, we demonstrate  a novel method to efficiently couple a single mode fiber to a dielectric nanophotonic waveguide using a conical tapered fiber tip. The coupling is based on an adiabatic transfer of the fundamental mode of the optical fiber to the fundamental mode of the nanophotonic waveguide. 
Our method can be applied to  general dielectric one-dimensional waveguides. In contrast to biconical tapered fibers \cite{groblacher13}, our devices are single-sided, thereby offering alternative geometries and mechanical support for nanophotonic devices \cite{thompson13a,tiecke14} and opening the door for new applications.

\section{Adiabatic coupling}

Adiabatic mode transformation is widely used to obtain efficient power transfer through nonuniform optical waveguides \cite{loveBook}. The key idea is to change the waveguide cross-section slowly along the propagation direction of the light such that all the optical power remains in a single eigenmode of the composite waveguide, while the coupling to other modes is suppressed. 
More specifically, two eigenmodes $\nu$ and $\mu$ with respective effective indices $n_\nu$ and $n_\mu$ define a characteristic beating length scale between the modes given by $z_b=\lambda/(n_\nu-n_\mu)$, where $\lambda$ is the wavelength in vacuum. In order to achieve adiabatic transfer  the characteristic length scale $z_t$ over which the waveguide changes has to be large compared to $z_b$ \cite{love91}. While the exact coupling strength depends on the details of the spatial mode profiles, we design our devices according to this intuitive length-scale argument and use numerical simulations to verify the design. 

Figure 1a shows a schematic of a typical device. A fiber is tapered down to a conical tip and an inversely tapered silicon nitride ($\mathrm{Si_3N_4}$) waveguide is attached over $7.5\,\mu$m. In 
what follows,  we separate the fiber tips in two regions (see Fig.1a): the adiabatic fiber-waveguide coupler (I) and the tapered fiber (II), separated by plane P at the start of the waveguide. In both regions we design our devices according to the adiabaticity criterion by ensuring $z_t>z_b$. In region I the optical modes of the fiber and waveguide are coupled through their evanescent fields and form a set of hybridized supermodes. $z_b$ is determined by the difference between the effective indices of the fundamental supermode and the higher order supermode  with the closest propagation constant (in this case radiation mode) while   $z_t$ is limited by the length of the coupling region.  In region II the fiber is tapered down from a standard single mode fiber, adiabatically transforming the core-guided HE11 mode to a cladding-guided HE11 mode. We ensure that the local taper angle  $\Omega(z)\equiv\rho(z)/z_t \ll \rho(z) /z_b$, where $\rho(z)$ is the fiber radius at position $z$ along the taper. In region II  $z_b$ is determined by the coupling of the fundamental fiber mode to the nearest higher-order mode. We follow the conventional design for biconical adiabatic fiber tapers where recently transmission efficiencies as high as $99.95\,\%$ \cite{hoffman14} have been achieved (see also \cite{love91,stiebeiner10} for details). 

For a tapered nanophotonic waveguide which vanishes at plane P (such as that shown in figure 1) the cross-section of refractive index profile changes smoothly along the complete coupler. For such waveguides the criteria given above are sufficient to design the coupler. However, for a waveguide with a non-vanishing cross-section at P (such as the rectangular waveguide shown in figure 1d), the refractive index cross-section changes discontinuously. We approximate the power transfer through P by projecting the fundamental fiber mode on the fundamental supermode  at P. In order to achieve efficient power transfer, we design the coupler such that this projection is close to unity. In this  case the effect of the waveguide is only a small perturbation of the fundamental fiber mode at plane P, thus enabling us to design couplers based on simple eigenmode calculations.

\begin{figure}[ht!]
\includegraphics[width=1.0\columnwidth]{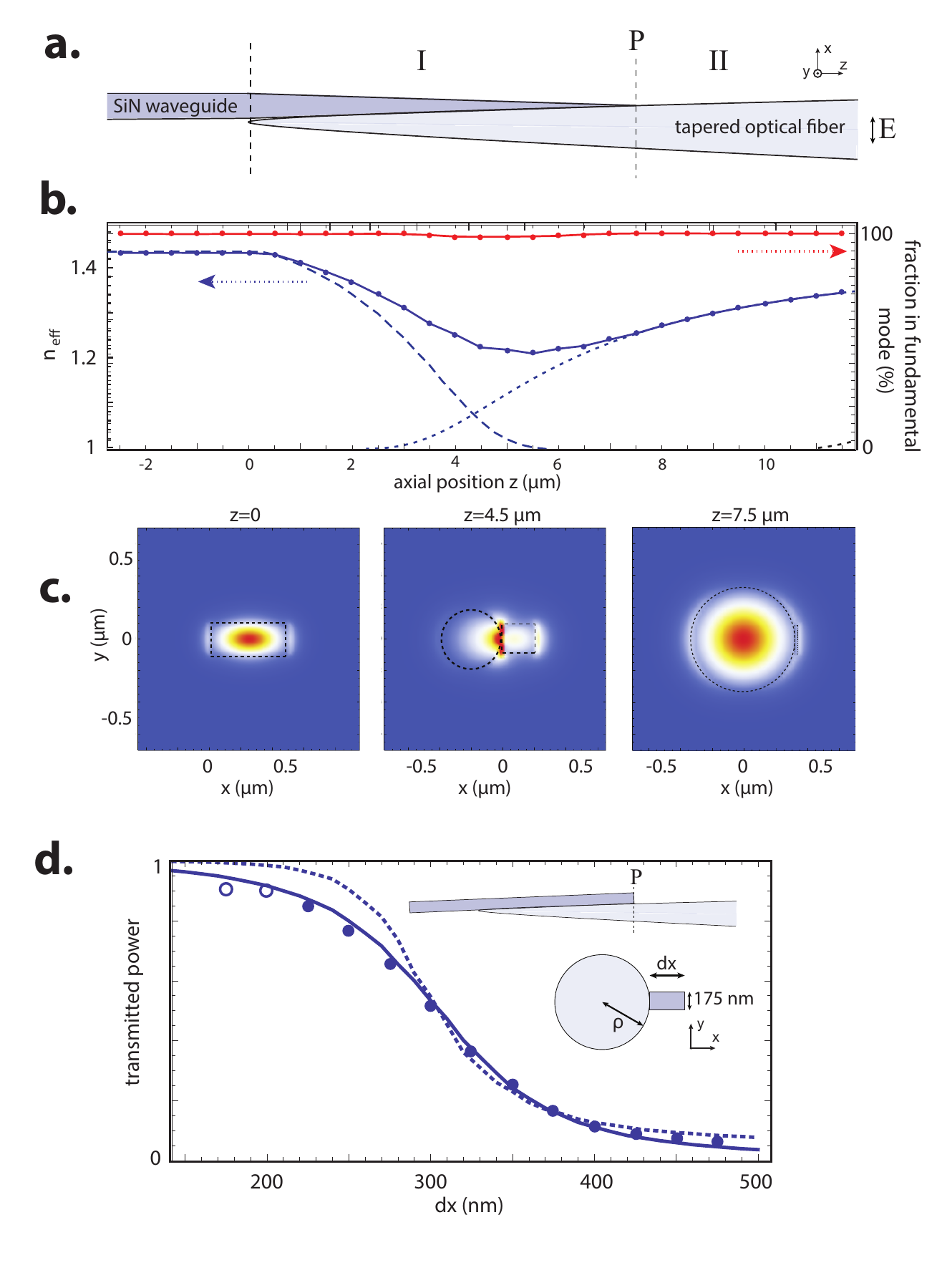}
    \caption{\textbf{Adiabatic transfer between fiber and waveguide modes.} \textbf{a.} Schematic of the fiber-waveguide coupling. The fiber (right) has a conical shape and is attached to a tapered $\mathrm{Si_3N_4}$ rectangular waveguide (left) and we consider modes polarized along $\hat x$. \textbf{b.} Effective index $n_\mathrm{eff}$ of the fiber and waveguide modes for an opening angle of the fiber (waveguide) of $5^\circ$ ($4^\circ$).  The blue dotted (dashed) lines are the separate fiber (waveguide) modes and the blue solid line corresponds to the fundamental supermode of the combined structure. The red line shows the power in the fundamental supermode obtained from an FDTD simulation of the coupler (see text). \textbf{c.} Cross sections of $|E|^2$ obtained from the FDTD simulation at various points along the coupler. \textbf{d.} The fraction of the power in the fundamental supermode of the combined structure as a function of the waveguide width $dx$, obtained from a mode decomposition (solid line). The transmission through a tapered coupler (see \emph{inset}) obtained with an FDTD simulation (circles) agrees well with the estimated transmission obtained from the mode decomposition. The two data points for $dx\leq 200\,\mathrm{nm}$ (open circles) are calculated using a shallower fiber angle ($2^\circ$) to ensure $z_t>z_b$. The dotted line shows the same geometry except that the fiber and waveguide are in contact on the $xz$-plane instead of the $yz$-plane. The fiber-waveguide cross-sections used for this simulation are shown in the \emph{inset},  $\rho=450\,\mathrm{nm}$.
    \label{fig:fig1}}
\end{figure}

\begin{figure}[t!]
\includegraphics[width=1.0\columnwidth]{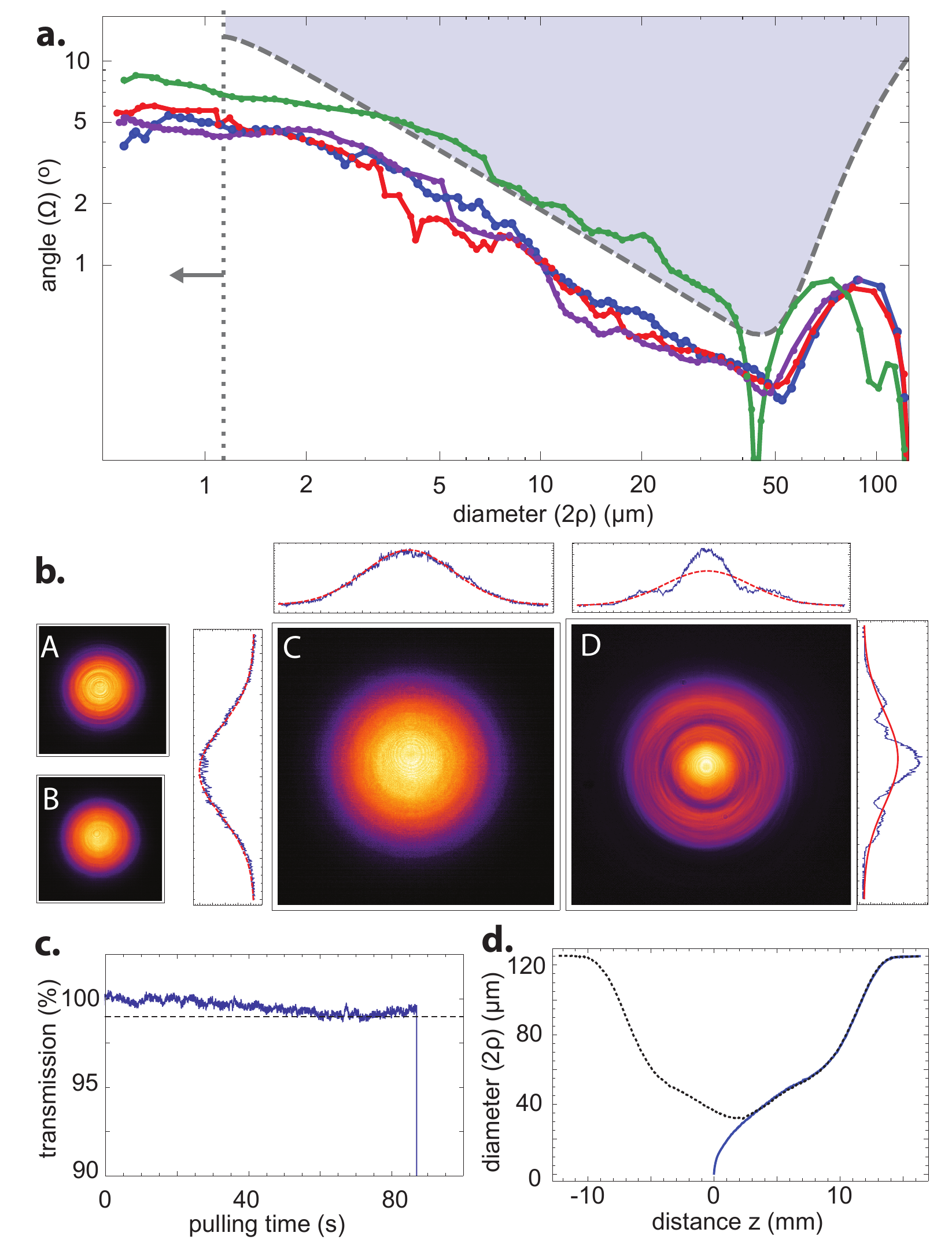}
    \caption{\textbf{Characterization of adiabatic tapers.} \textbf{a.} The fiber angle as a function of the local fiber diameter along the taper axis $z$. The dashed line and shaded area indicate the adiabaticity criterion $z_t>z_b$ as discussed in the text. 
    Fiber tapers which have a profile below the dotted line are expected to be adiabatic. For a diameter smaller than $1.1\,\mu \mathrm{m}$ the HE12 mode is cut off. The taper profiles for 4 tapers (blue (A), red (B), purple (C) and green (D)) are shown. \textbf{b.} Far field mode profiles. Tapers A, B and C show Gaussian profiles, while taper D has clear contributions from higher order modes. For tapers C and D cuts through the center of the profiles are shown together with a Gaussian fit. \textbf{c.} The transmission versus pulling time of a taper similar to A-C, the dashed line indicates $99\%$ transmission. The sudden drop in transmission at $\simeq 87\,\mathrm{s}$ arises from the fast pull by the electromagnetic coil. \textbf{d.} The taper profile of taper C (blue) and of a biconical taper (dashed) using the same pulling parameters but without pulsing the electromagnet to create the tip.
   \label{fig:fig2}}
\end{figure}

We next verify these design criteria using Finite Difference Time Domain (FDTD) simulations. Figures 1b and 1c show simulations for the device presented in figure 1a. We consider a conical fiber with an opening angle $d(2\rho)/dz=5^\circ$ and a $dy=175\,\mathrm{nm}$ thick $\mathrm{Si_3N_4}$ waveguide with a taper angle $d(dx)/dz=4^\circ$ to a width of $dx=500\,\mathrm{nm}$. We focus on TE-polarized ($\hat x$) modes, but we have verified that the same reasoning can be applied for TM-polarized modes. Figure 1b shows the effective indices of the fiber mode, the waveguide mode and of the hybridized mode of the combined structure (supermode). For this geometry, the latter has an effective index of $n_{\mathrm{eff}}>1.2$ over the entire length of the coupler. The combined structure supports only one other mode, which, however, has orthogonal polarization and therefore does not couple to the fundamental supermode.
The relevant beat length is therefore set by the fundamental supermode and the free space modes ($n_0 = 1$), and corresponds to $z_b\simeq 4\,\mu  \mathrm{m}$. We chose the length of the coupler ($z_t \simeq 7\,\mu \mathrm{m}$) to be longer than $z_b$. 
In order to verify adiabaticity we perform a FDTD simulation (see figure 1b, c), in which we excite the fiber taper at $z=11\,\mu \mathrm{m}$ with the fundamental HE11 mode polarized along $\hat x$ and propagating along $-\hat z$. Along the coupler we decompose the optical fields in the basis of local eigenmodes of the combined fiber-waveguide structure and find that essentially all the optical power ($>99\,\%$) is in the fundamental mode across the complete
fiber-waveguide coupler, thereby confirming the adiabaticity of the coupler. 

In the case of a rectangular waveguide (Figure 1d), we  model the sudden onset of the waveguide by decomposing the fundamental fiber mode in the basis of supermodes of the combined structure. This decomposition is performed using the fields of the eigenmodes of the fiber and the combined structure which we obtain using the MIT Photonic Bands (MPB) mode-solver \cite{johnson01}. To verify that the projection indeed describes the power transfer accurately we compare the mode decomposition results with FDTD simulations. Figure 1d shows the power in the fundamental supermode of the combined structure as obtained from the mode decomposition and from an FDTD simulation for a rectangular $\mathrm{Si_3N_4}$ waveguide with varying width $dx$ and a fiber with radius $\rho=450\,\mathrm{nm}$. We find that the loss of transmission through the coupler obtained from the FDTD simulation is well described by the mode decomposition. For the simulated conditions the losses can be made small for waveguide dimensions below 200 nm.   

\section{Design and fabrication}

We next discuss the optimization and characterization of the tapered fiber tips in region II. For our experiments we use single mode fiber (Thorlabs 780HP), with a $4.4\,\mu \mathrm{m}$ ($125\,\mu \mathrm{m}$) core (cladding) diameter and we optimize our design for  a wavelength of $\lambda=780\,\mathrm{nm}$. Figure 2a shows the critical angle ($\Omega_c(z)=\rho(z) /z_b(z)$) for this fiber. In a cylindrically symmetric geometry, modes with different angular momentum do not couple, therefore, the coupling occurs between the HE11 and HE12 modes. At large fiber-diameters ($d> 50\,\mu \mathrm{m}$) the adiabaticity criterion is determined by coupling of the HE11 core guided mode and the cladding guided modes, while for $d<50\,\mu \mathrm{m}$ the adiabaticity criterion is determined by the coupling of the HE11 and HE12 cladding guided modes. 

We fabricate fiber tapers using a conventional heat-and-pull setup \cite{mazur03,ward14} where the fiber is heated using an isobutane torch ($140\,\mathrm{mL/min}$ flow), with an effective flame length of $L=4.3\,\mathrm{mm}$. The flame is continuously brushed back and forth to heat the fiber over a variable length, which is adjusted during the pulling to obtain the desired fiber profile (see Refs. \cite{mazur03,bambrilla10}). This results in a $24\,\mathrm{mm}$ long biconical fiber taper with a minimum diameter of $\sim\,30\mu \mathrm{m}$. At this stage we apply a fast pull to one of the stages holding the fiber, which quickly ($\sim 10\,\mathrm{ms}$) pulls the fiber out of the flame thereby creating a $14\,\mathrm{mm}$ long fiber taper with a conical tip (see figure 2d). 
The fast pull is generated by an electromagnet, composed of a hard-drive head with its arm connected to one of the two fiber clamps. 
The clamp itself is mounted on a linear ball-bearing translation stage and a current pulse through the electromagnet results in a constant acceleration of the fiber, creating a smooth fiber tip which is well described by a parabolic shape at larger fiber diameters and a constant opening angle over the last tens of microns. We find that the acceleration changes linearly with the applied current over a range of $17\,\mathrm{m/s^2}$ to $46\,\mathrm{m/s^2}$ and we typically use an acceleration of $33\,\mathrm{m/s^2}$. By optimizing the heat-and-pull-parameters we realize the requirements of the taper angle for large diameters, while the electromagnet current and fiber-diameter at which the pulse is applied controls the fiber taper angle at smaller diameter. We note that the resulting parabolic shape of the fiber taper conveniently has the same scaling ($\Omega\sim1/\rho$) as the adiabaticity criterion at the relevant range of fiber diameters ($2 - 50\,\mu \mathrm{m}$, see figure 2a). Additionally, since our fiber tips have sub-wavelength dimensions only over $\sim 10\,\mu \mathrm{m}$, the requirements on the cleanliness of the flame and the fabrication environment are less stringent as compared to those for creating efficient biconical tapered fibers \cite{hoffman14,ward14}.

\section{Characterization} 

We characterize our devices with several measurements. First, we measure the taper profiles to ensure the local angle is smaller than the critical angle set by the adiabaticity condition. Figure 2 shows three fibers (A, B, C) which are made under the same conditions, while fiber D is made using different pulling parameters for the purpose of illustrating the performance of a sub-optimal fiber taper. In figure 2a we show the fiber profiles for each fiber, which are measured using optical and scanning electron microscopy (SEM). Fibers A, B and C show nearly identical profiles which satisfy the adiabaticity criterion, indicating that our fabrication method yields reproducible fiber tapers. Fiber D has a somewhat steeper angle for fiber diameters around $\sim 3-30\,\mu \mathrm{m}$ and is therefore expected to be non-adiabatic. Second, in figure 2b we show the far-field profiles of the fiber mode imaged directly on a CMOS camera. The optical modes of fibers A, B, C are all nearly Gaussian, indicating that at the end of the taper, most of the power is in the fundamental HE11 mode. Fiber D shows clearly a multimode structure, in agreement with our expectation that this taper does not match the adiabaticity criterion. To quantify the single mode character of the profiles we calculate the coefficient of determination ($R^2$) of the Gaussian fits, resulting in $R^2=0.98,0.99,0.99,0.87$ for tapers A, B, C and D, respectively. We find this method of measuring the far field profiles to be a very fast, sensitive and reliable for verifying  the single mode character of our fiber tips.
Third, in figure 2c we show the total transmission during the time of the pulling-process. We observe $>99\%$ transmission during the complete pulling process; consistent with the full biconical fiber taper being single-mode before the fast pull occurs.
 
After confirming that our fiber tips are single mode, we measure the coupling efficiency ($\eta_c$) to a tapered $\mathrm{Si_3N_4}$ waveguide (see Refs. \cite{thompson13a,tiecke14} for details of the device fabrication). 
The waveguide we use has a photonic crystal cavity, which, away from the cavity resonance, reflects all the incident light. We measure the reflected power $P_\mathrm{r}$, normalized to the power $P_\mathrm{in}$ in the fiber before the taper is pulled (see Fig. 3a). The normalized reflection is given by $P_\mathrm{r}/P_\mathrm{in}=\eta_c^2 \eta_m \eta_{bs} \eta_{FC}$, where $\eta_{bs}$ and $\eta_{FC}$ are the fiber beamsplitter and FC-FC coupling efficiencies and $\eta_m$ is the Bragg mirror reflectivity (see SI). We obtain a coupling efficiency of $\eta_c=0.97(1)$, where the error bars reflect drifts of the input power and calibrations over the course of our measurements.
In comparison we achieve coupling efficiencies of $\eta_c\simeq 0.5 - 0.6$ for fiber tips with parameters such as fiber D.  
Figure 3d shows measurements of the coupling efficiency for waveguide angles of $2^\circ \leq \alpha \leq 8^\circ$ and for rectangular waveguides with a width ranging from: $100 <dx< 250\,\mathrm{nm}$, all waveguides have a $dy=175\,\mathrm{nm}$ thickness. For these measurements the  waveguides are attached to the chip on one side only (see figure 3b). We observe that the coupling efficiency for most of these devices is $\geq 95\,\%$. The coupling efficiencies for $dx\leq 150\,\mathrm{nm}$ are slightly lower, consistent with a not fully adiabatic coupler ($z_b>z_t$) since for decreasing waveguide width $z_b$ increases while in our measurement we keep $z_t$ constant. 

\begin{figure}[ht!]
\includegraphics[width=1.0\columnwidth]{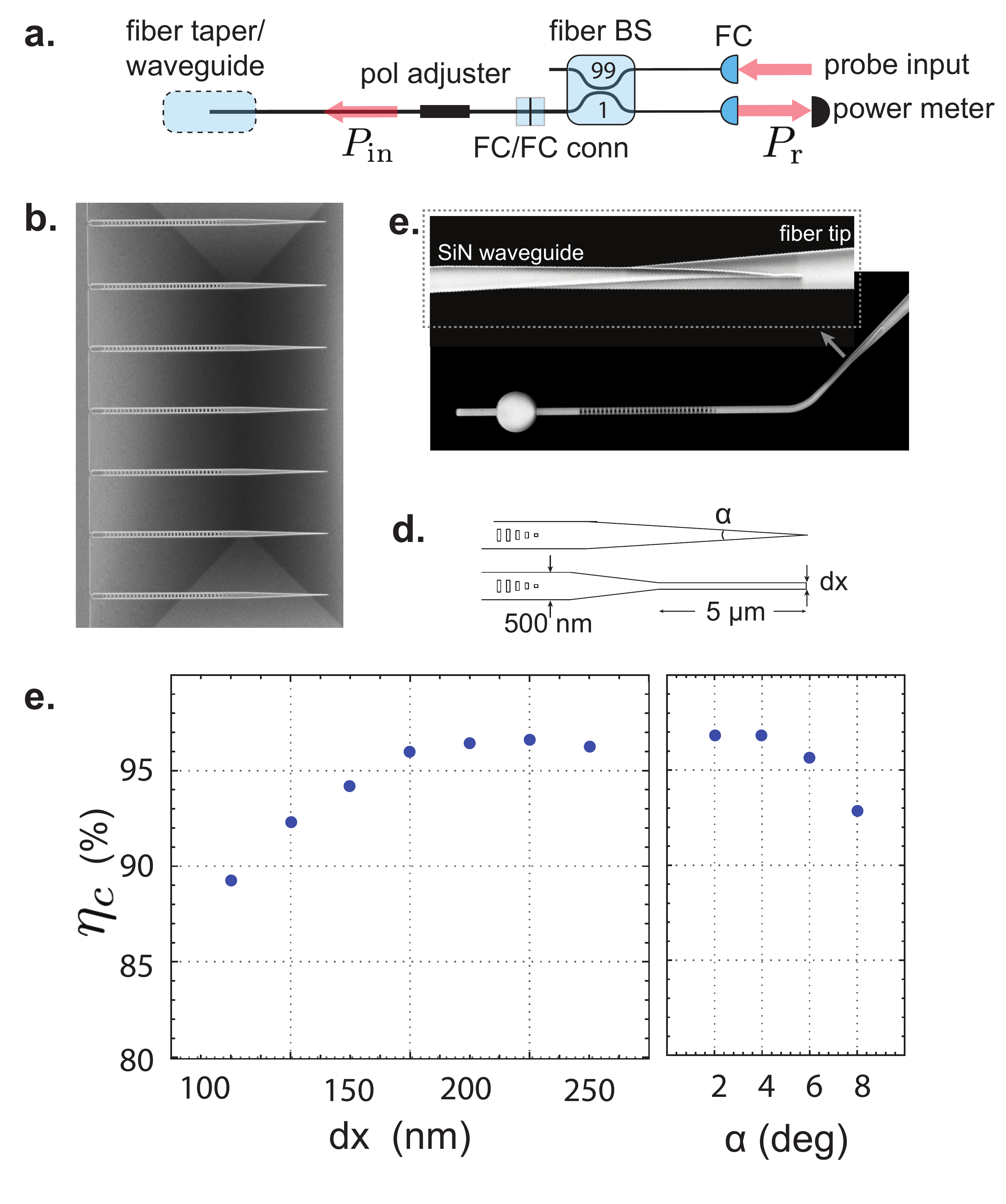}
    \caption{\textbf{Coupling to photonic crystal waveguide cavities.} \textbf{a.} Setup to measure fiber-waveguide coupling efficiency. A tunable probe laser is coupled weakly to the fiber connecting to the device using a 99:1 fiber beamsplitter. The polarization at the waveguide is adjusted by means of a fiber polarization controller and the light is in and out coupled of the fiber network using fiber collimators (FC) \textbf{b.} SEM image of an array of singly-clamped photonic crystal waveguide cavities used for on-chip measurements. 
\textbf{c.} SEM image of a photonic crystal cavity attached to the fiber tip, \emph{inset} shows a zoom of the fiber-waveguide coupler.  \textbf{d.} Schematic of the various waveguide geometries. \textbf{e.} Coupling efficiencies for a range of waveguides; the devices are either a tapered waveguide with an opening angle $\alpha$ or rectangular waveguides with a varying width $dx$ and $5\,\mu \mathrm{m}$ long before adiabatically expanding to the photonic crystal cavity. All waveguides are $175\,\mathrm{nm}$ thick and attached to the chip as in panel \textbf{b}.   }%
   \label{fig:fig3}
\end{figure}

We also detach cavities from the chip and attach them to the fiber such that they are solely connected to the fiber tip (see SI). A typical device attached in free space is shown in figure 3c for which we measure a coupling efficiency of $\eta_c=0.96(1)$.
We find that our alignment procedure (see SI) allows to optimize the coupling efficiency in a reliable and reproducible manner, however, since we perform the alignment under an optical microscope we do not have exact knowledge of the fiber-waveguide interface. From our simulations we find that for a range of configurations the coupling efficiency is close to unity and consistent with the design criteria defined above. 

\section{Outlook}

We have presented a method for highly efficient fiber coupling to nanophotonic waveguides. Our measurements indicate  coupling efficiencies as high as $97(1)\%$ for a range of devices. These results open the door for a range of unique applications in quantum optics and nano photonics. In particular, in combination with our recent results demonstrating strong coupling of a single atom to photonic crystals \cite{thompson13a,tiecke14}, efficient coupling to fibers can enable the creation of highly non-classical Schr\"odinger cat states of light \cite{wang05}  and realization of efficient protocols for scalable quantum networks \cite{kimble08}. Moreover, the flexible geometries as well as the fiber-based mechanical support for nanophotonic devices, allowed by this approach  open the door for  new applications in nanoscale biosensing \cite{tan92,yan12,shambat13}.

\section*{Funding Information}
Financial support was provided by the NSF, the Center for Ultracold Atoms, the Natural Sciences and Engineering Research Council of Canada, the Air Force Office of Scientific Research Multidisciplinary University Research Initiative and the Packard Foundation. KPN acknowledges support from Strategic Innovation Program of Japan Science and Technology Agency (JST). JDT acknowledges support from the Fannie and John Hertz Foundation and the NSF Graduate Research Fellowship Program. This work was performed in part at the Center for Nanoscale Systems (CNS), a member of the National Nanotechnology Infrastructure Network (NNIN), which is supported by the National Science Foundation under NSF award no. ECS-0335765. CNS is part of Harvard University.

\setcounter{figure}{0} \renewcommand{\thefigure}{S\arabic{figure}}

\section*{Supplementary Information}

For all measurements we assume perfect reflection from the Bragg mirror ($\eta_m\equiv 1$) and we correct for the independently calibrated beamsplitter ratio ($\eta_{bs}=0.99$) and the FC-FC fiber coupling (with a typical value of $\eta_{FC}=0.89$).
We note that our results for $\eta_c$ are conservative since we observe a small amount of scattering from the entrance Bragg mirror ($\eta_m<1$) and did not account for propagation losses through the $\simeq 5\,\mathrm{m}$ of fiber. We estimate these combined losses to be on the percent level, however, our current method has not sufficient accuracy to determine $\eta_c$ to a higher precision. 

Aside from the high coupling efficiency another important property of the coupler are its possible reflections. If light is reflected from the fiber-waveguide interface rather than the Bragg mirror this could affect measurements performed in reflection, such as the fiber-wg coupling efficiency characterization performed here, or quantum optics experiments.
In order to characterize spurious reflections from our coupler we perform a measurement using a critically-coupled
double-sided cavity. In the absence of cavity losses the reflection vanishes on resonance (enabling us to measure spurious reflections from the coupler) and the cavity is fully reflective off resonance (enabling us to measure the fiber-wg coupling efficiency as described above and in the main text). We choose a low quality factor ($Q\simeq 2000$) in order to minimize the effect of cavity losses and detach the cavity from the chip to avoid additional reflections from the chip. For this specific device we achieve a slightly sub-optimal coupling efficiency of $\eta_c=0.87$ and for TE-polarized light (see figure S1) we observe an on-resonance reflection of $2.0(4)\times10^{-3}$. Additionally, we do not observe any reflection ($0.1(3)\times10^{-3}$) of the fiber tip when no cavity is attached. Finally, in an independent measurement using TM-polarized light we measure a reflectivity of $4(1)\times10^{-3}$. These values set an upper limit of the reflection from the coupler since we have assumed a perfectly symmetric cavity, no cavity losses, perfect polarized probe light and no reflection of TM-polarized light by the Bragg mirror. Here, we have assigned the polarizations based on maximum and minimum reflection from the Bragg mirror which is designed to have a band gap for TE-polarized light at 780 nm. 

In order to align the relative fiber-waveguide position we mount the fiber under an angle of $\sim 10-20^\circ$ and align the fiber with respect to the waveguides under an optical microscope using a three axis translation stage with micrometer and piezo control. When the fiber is brought into contact with the waveguide, they stick together, allowing stable alignment to be maintained over long periods of time. They can be released from each other by pulling them apart with the translation stage, apparently without damage. While we have not investigated the mechanism of the sticking, we note that "stiction" is very common and has been extensively studied in the context of micro- and nano-mechanical systems.

We obtain the optimal fiber-waveguide coupling by adjusting the fiber position, optimizing for the reflected power. Typically, the optimal coupling is achieved by slightly lifting the fiber off the chip which we attribute to the waveguide bending and aligning with the angle-mounted fiber taper. 
The procedure to detach cavities from the chip and attach them to the fiber taper is as follows: we break off and pick up a waveguide using a tungsten tip and transfer  it to the fiber tip in free space using three axis translation mounts for both the fiber and tungsten tip. We move the waveguide using the tungsten tip while optimizing for optimal reflection. 

\begin{figure}[t]
\includegraphics[width=1.0\columnwidth]{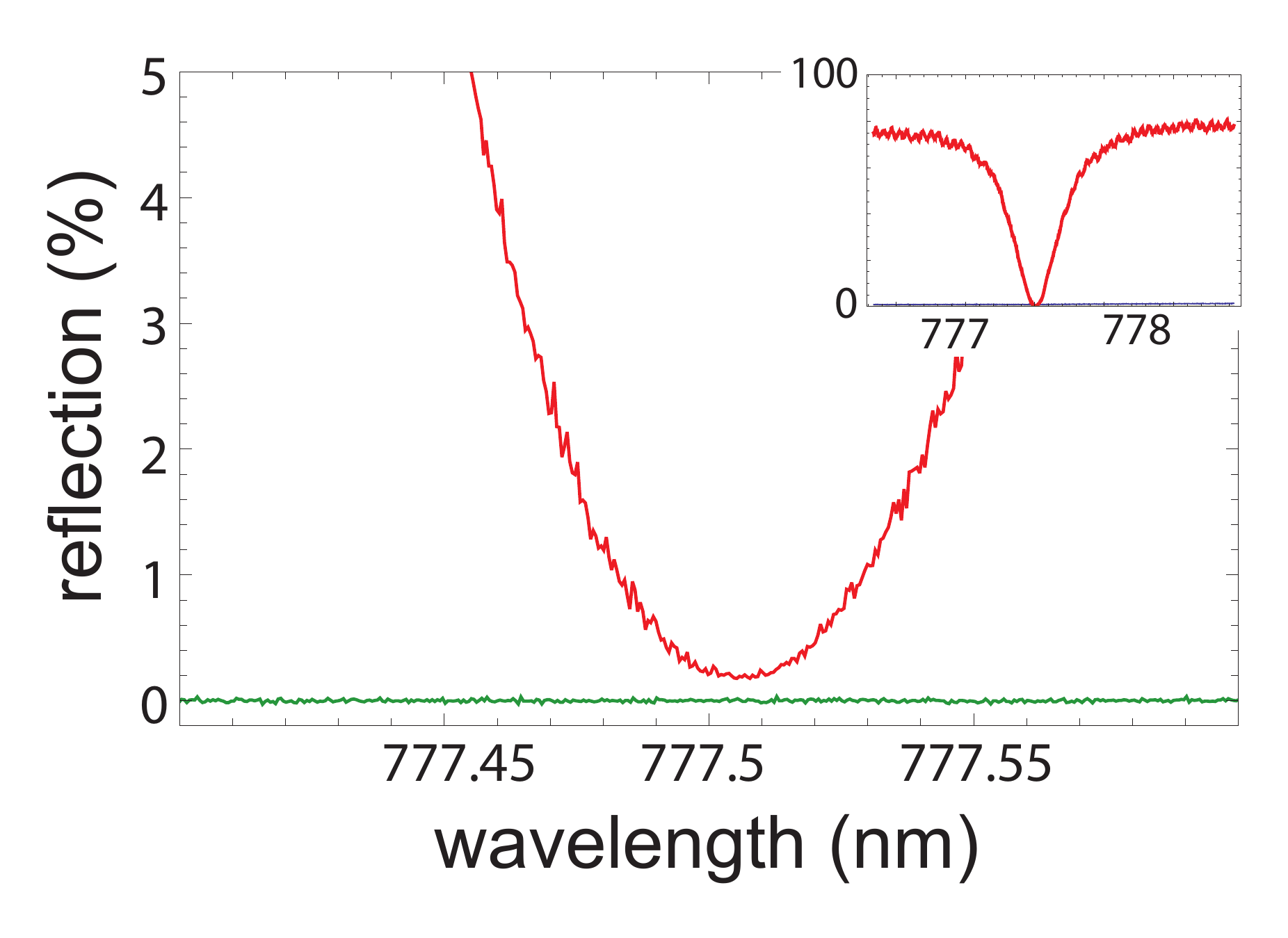}
    \caption{\textbf{Spurious reflections from coupler.} Reflection spectrum of a critically damped cavity (\emph{Inset} shows full spectrum) for TE-polarized light (red) and for the bare fiber taper without a cavity attached (green).}%
   \label{fig:fig3}
\end{figure}

\end{document}